\documentstyle[prl,aps,twocolumn]{revtex}
\begin{document}

{\bf K\"uhn and Mazzeo Reply} 
Van Enter, K\"ulske and Maes (EKM) in their comment\cite {ekm}
to our letter\cite{ku94} point out -- correctly -- that joint 
probability measures on the product of spin- and disorder-space in 
systems with quenched randomness are non--Gibbsian. They conclude 
from this  -- incorrectly in view of the connotations carried by 
such a characterization -- that our approach which provides 
approximate descriptions of such systems in terms of a net of 
Gibbs measures suffers from being ill defined.

To wit, it has been shown\cite{ku97} that the approximation scheme
used in \cite{ku94} provides a {\em variational family of increasing 
exact lower bounds\/} for the free energy of systems whith quenched 
randomness in terms of Gibbs measures on the joint spin- and 
disorder-space. These measures agree on an increasing set of moments 
with the measure characterizing the quenched disorder. Moreover, at 
each level of the approximation studied in \cite{ku94,maku99} 
the variational problem is convex (also in the thermodynamic limit), 
hence its solution unique. In this sense, our approximation scheme 
is perfectly well defined, and it is legitimate to investigate its 
efficiency in analyzing unsolved problems in the physics of systems 
with quenched randomness.

The joint measure related to the 2$D$ spin-diluted Ising model 
studied in \cite{ku94,maku99} is Gibbsian in the high-temperature 
phase, while only weakly Gibbsian in the ferromagnetic low-temperature 
phases (for definitions see the literature cited in \cite{ekm}). 
Given the representation of the approximate measures studied in \cite
{ku94,maku99}, it is perfectly conceivable that they could approach the 
joint measure of the quenched system, if one were able to carry the 
approximative scheme to its end. While a formal proof of this statement 
may be difficult, it nonetheless indicates that differences between 
the limiting measure of our scheme and the weakly Gibbsian measure 
it attempts to approximate in the low temperature phases might be 
rather subtle. 

Rather than formal convergence properties, which ultimately
require clarification, our main issue from a pragmatic point 
of view is the {\em efficiency\/} of the approximations in
describing physical properties. As is quite common, we are
only able to work through the first few steps of the scheme,
hence the pragmatic question: Is the physics we are thereby 
missing essential or not? As far as one can presently see, it 
is not. Our findings concerning critical phenomena in 
\cite{ku94,maku99} are fully in line with currently available 
empirical evidence coming from other conventional approaches, 
as well as with known exact results. The fact that critical 
behavior is, nevertheless, still under debate for the 2$D$ 
spin-diluted Ising model is related to the difficulty in disentangling 
the finite size signatures of the conflicting scenarios from the 
numerical data, irrespective of what kind of (efficient) 
approximation is being used to obtain them\cite{maku99}. 

The physics we {\em are\/} likely to miss generally is that 
associated with Griffiths' singularities \cite{gri69}. We have 
been able to prove this for the 1$D$ system \cite{ku97}: the 
simplest approximation in this case (system (a) of \cite{ku94}) 
is exact in zero field $h=0$, and it also very accurately describes 
critical behavior and scaling in the vicinity of the multicritical 
point $T=h=0$, $\rho=1$, but fails to exhibit Griffiths singularities. 
The absence of these very weak essential singularities, while certainly 
a drawback, does in the light of available evidence not appear to be 
a very serious one for the problem studied in \cite{ku94,maku99}.

The important message of EKM is that formal convergence issues of 
our approach have indeed not yet been dealt with in a satisfactory 
way. A number of such issues occurring in situations where 
non-Gibbsian measures do play a potentially worrying role have been 
studied before, mainly in the renormalization group (RG) context 
(see in particular \cite{efs}, and other references in \cite{ekm}). 

However, our case is also different from the RG setting in essential 
aspects, the perhaps deepest one being that the disorder potentials
used to express approximating measures are invariant under {\em local\/} 
spin-reversal, a feature which may well remove many of the pathology 
generating mechanisms occuring in the RG context. 

In summary, there is clearly nothing automatic about the success of 
our approach, nor, however, about its potential failure due to some 
formal similarities with the RG situation. Independent evidence must in 
any new case be checked. We have done no less in \cite{ku94,maku99}, 
with clearly encouraging results.

We are indebted to Charles Newman for very useful discussions. G.M. 
acknowledges support from the Foundation Blanceflor Boncompagni-Ludovisi.

\bigskip\noindent
{\small Reimer K\"uhn${}^{(a)}$ and Giorgio Mazzeo${}^{(a,b)}$\\
${}^{(a)}$Institut f\"ur Theoretische Physik, Universit\"at 
Heidelberg\\~~Philosophenweg 19, 69120 Heidelberg, Germany\\
${}^{(b)}$ Courant Institute of Mathematical Sciences\\~~New York 
University, New York, NY 10012-1185, USA \\
and Dipartimento di Fisica, Universit\`a di Genova \\
via Dodecaneso 33, 16146 Genova, Italy}

\date{Jan 18 2000}

	\pacs{PACS: 75.10Mg 75.10Hk}

\end{document}